\documentclass{article}

\usepackage[nonatbib,final]{neurips_2019}

\usepackage[utf8]{inputenc} 
\usepackage[T1]{fontenc}    
\usepackage{hyperref}       
\usepackage{url}            
\usepackage{booktabs}       
\usepackage{amsfonts}       
\usepackage{nicefrac}       
\usepackage{microtype}      
\usepackage{graphicx}
\usepackage{multirow}
\usepackage{capt-of}
\usepackage{todonotes}
\usepackage[normalem]{ulem}

\title{Variational Inference and Bayesian CNNs for \\Uncertainty Estimation in \\ Multi-Factorial Bone Age Prediction}

\author{%
	Stefan Eggenreich\\
	Institute of Computer Graphics and Vision\\
	Graz University of Technology, Austria \\
	\texttt{stefan.eggenreich@student.tugraz.at} \\
	\And
	Christian Payer \\
	Institute of Computer Graphics and Vision\\
	Graz University of Technology, Austria \\
	\texttt{christian.payer@icg.tugraz.at} \\
	\And
	Martin Urschler \\
	School of Computer Science \\
	University of Auckland, New Zealand \\
	\texttt{martin.urschler@auckland.ac.nz} \\
	\And
	Darko \v{S}tern \thanks{This work was supported by the Austrian Research Promotion Agency (FFG): 871262, and partly funded by the City of Graz, Kulturamt,  GZ: A 16 – 021628/2013.}\\
	Institute of Computer Graphics and Vision\\
	Graz University of Technology, Austria \\
	\texttt{stern@icg.tugraz.at}
}

\begin{document}

\maketitle


\begin{abstract}
    Additionally to the extensive use in clinical medicine, biological age (BA) in legal medicine is used to assess unknown chronological age (CA) in applications where identification documents are not available.
    Automatic methods for age estimation proposed in the literature are predicting point estimates, which can be misleading without the quantification of predictive uncertainty.
    In our multi-factorial age estimation method from MRI data, we used the Variational Inference approach to estimate the uncertainty of a Bayesian CNN model.
	Distinguishing model uncertainty from data uncertainty, we interpreted data uncertainty as biological variation, i.e. the range of possible CA of subjects having the same BA.

\end{abstract}

\section{Motivation}

Estimation of the progress of physical maturation in young individuals, i.e. biological age (BA) estimation, is extensively used in clinical medicine, e.g. for diagnosis of endocrinological diseases. 
In legal medicine, BA estimation is used as an approximation to assess unknown chronological age (CA) in applications where identification documents of individuals are not available. 
  
Established radiological methods for estimation of BA are based on visual examination of the ossification of hand bones from X-ray images~\cite{greulich1959}.
Maturation progress is not simultaneous for all bones and by including radius and ulna into examinations it can be followed up to an age of around 18 years.
To extend this age range up to around 24 years, which is highly relevant in legal medicine applications, the workgroup for forensic age diagnostics (AGFAD) \cite{schmeling2011FAEUnaccompaniedMinors} recommended combining age-related information from hand bones in X-ray images with clavicle bones in chest computed tomography (CT) images and teeth in dental panoramic X-ray images.
Recent research in such multi-factorial age estimation proposes magnetic resonance imaging (MRI) for age estimation, to prevent the use of ionizing radiation.
To eliminate the need for defining discrete staging schemes for individual anatomical sites and subjective schemes for their fusion into an age estimate, automatic methods for age estimation
have already been developed~\cite{Thodberg2009BoneXpert,Spampinato2017,Stern2019,Larson2017}.
A recently proposed CNN based method for multi-factorial age estimation~\cite{Stern2018} provides a solution for automatic fusion of developmental information from 3D MRI scans of hand, clavicle, and teeth into a BA prediction.
Although some of the above mentioned automatic methods showed favorable accuracy compared with human performance, they have in common that they predict point estimates without providing information on the uncertainty of the age prediction. 
Especially in decision-making processes, where a prediction with high uncertainty is considered equally or even more misleading compared to an incorrect prediction, quantification of predictive uncertainty is crucial.

In this work, we propose a Bayesian CNN model to estimate a subject's age from multi-factorial MRI data, where we use a Variational Inference (VI) approach to estimate the uncertainty of this model. 
Differently to previous Bayesian methods based on VI, we are distinguishing model uncertainty from data uncertainty. 
Interestingly, data uncertainty in age estimation can be interpreted as biological variation, i.e. the range of possible CAs of subjects that have the same BA.

\section{Method}

In a Bayesian framework, we can differentiate two main types of uncertainty: epistemic uncertainty that captures uncertainty of model weights ($\omega$) and aleatoric uncertainty that captures noise inherent in the data ($\mathcal{D}$)~\cite{Kendall2017}.
While standard CNN models are not able to capture uncertainty, Bayesian CNNs with distributions placed over the weights not only offer robustness to overfitting, but also means for recovering both types of uncertainty. 
We use the Variational Inference method Bayes by Backprop~\cite{blundell15} to approximate the intractable true posterior probability distribution of the network weights $p(\omega \mid \mathcal{D})$ with the tractable variational distribution $q(\omega \mid \theta)$ by learning the variational parameters $\theta$ that minimize the following loss function:
\begin{equation}
\mathcal{F}(\mathcal{D},\theta) \approx \sum_{i=1}^{n} \log q(\omega^{(i)} \mid \theta)- \log p(\omega^{(i)})- \log p(\mathcal{D} \mid \omega^{(i)}),
\label{bayes_by_backprop}
\end{equation}
where $\omega^{(i)}$ is one set of sampled weights, $\theta$ are the parameters of the distributions from where the weights are being sampled, and $n$ is the number of sampled weight sets.
We assign $n=1$ as proposed by~\cite{Graves2011}.
To differentiate aleatoric from epistemic uncertainty, 
we model $p(\mathcal{D} \mid \omega^{(i)})$ with the Gaussian distribution $\mathcal{N}(y \mid \mu, \sigma^2)$, where $y$ is the CA of a subject, and $\mu$ and $\sigma^2$ are outputs of the network representing the Gaussian output layer.
We compare our Bayesian Convolutional Neural Network that learns both~$\mu$ and~$\sigma^2$ (BCNN with $\sigma$) with a plain BCNN that learns only~$\mu$.
Thus, the plain BCNN does not explicitly model aleatoric uncertainty.
Furthermore, we compare the BCNNs with a CNN that models solely the aleatoric uncertainty with a Gaussian output layer (CNN with $\sigma$), as well as a plain CNN that does not model any kind of uncertainties.

\paragraph{Experimental setup:}
We evaluated our method on a dataset of T1-weighted 3D MR images of wrists and clavicles from $N = 328$ subjects with a known CA approximately uniformly distributed between 13.0 and 25.0 years.
In contrast to~\cite{Stern2018}, we do not use 3D volumes as network input, but the 2D middle slices of cropped radius and ulna (i.e. wrist) bones and both clavicle bones.
We utilized the same network architecture and data augmentation as proposed in~\cite{Stern2018}, but modified for 2D. 
For the BCNN architecture both convolutional and dense layers were replaced by their probabilistic counterparts (Reparameterization layers)~\cite{Kingma2014b}.
The final fully connected layer was modified to generate two regression outputs ($\mu$, $\sigma^2$).
During inference, weights were sampled from their posterior distribution in each forward pass. 
BA ($\hat{\mu}$) and data uncertainty ($\hat{\sigma}^2$) were estimated by averaging their predictions over $20$ forward passes of the test image.
For optimizing our models, we used Tensorflow Probability~\cite{Dillon2017}.
In our experiments we evaluated the performance using a 70:30 train-test split such that the age distribution was as close as possible in both training and test set. 
{The experiments were performed for wrist and clavicle bones combined, as well as for wrist bones alone.}
 
\section{Results and Discussion}

\newcommand{\width}{0.27\textwidth}

\begin{figure}
	\centering
	\begin{tabular}{ccccc}
		& & CNN with $\sigma$ & BCNN & BCNN with $\sigma$ \\
		\multirow{3}{*}{\rotatebox{90}{\makebox[5cm]{\small Trained on Wrist and Clavicle Bones}}} &
		\rotatebox{90}{\makebox[0em]{~~~~~\tiny{Groundtruth Age}}} &
		\begin{minipage}{\width}
			\includegraphics[width=\linewidth]{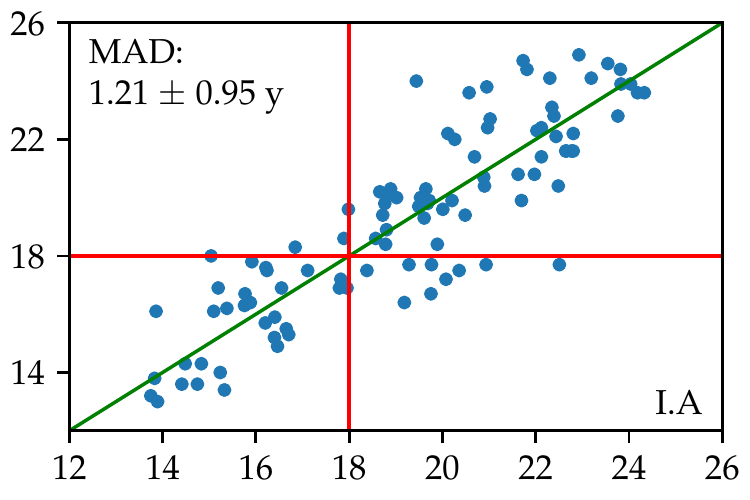}
		\end{minipage} &
		\begin{minipage}{\width}
			\includegraphics[width=\linewidth]{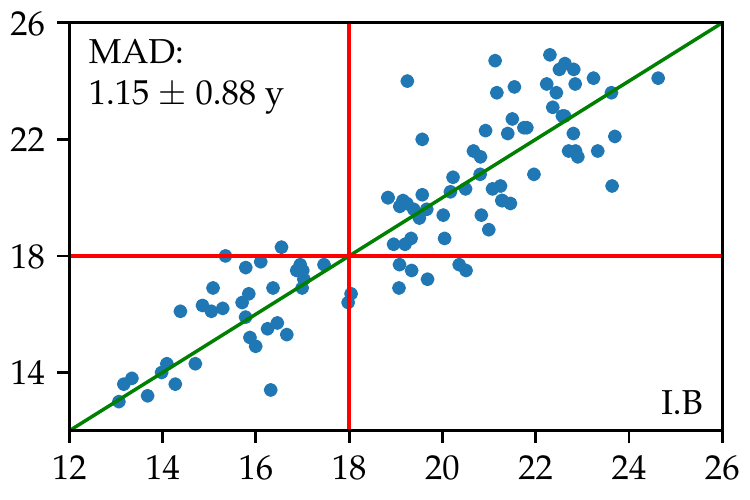}
		\end{minipage} &
		\begin{minipage}{\width}
			\includegraphics[width=\linewidth]{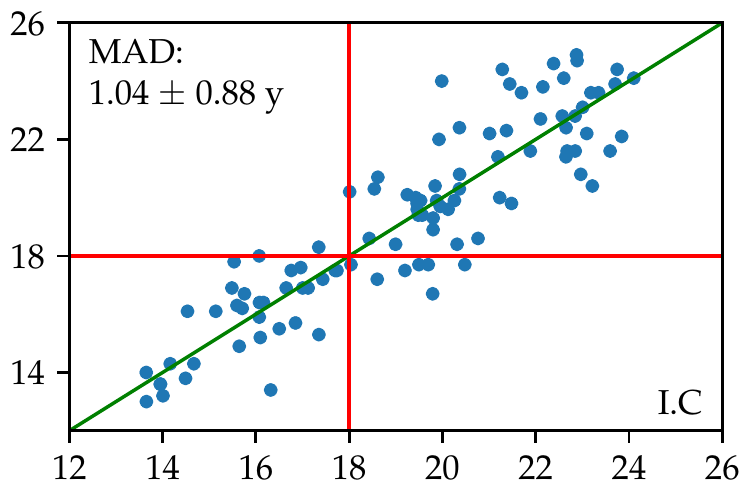}
		\end{minipage}\\
		& \rotatebox{90}{\makebox[0em]{~~~~~\tiny{Epistemic Uncertainty}}} & &
		\begin{minipage}{\width}
			\includegraphics[width=\linewidth]{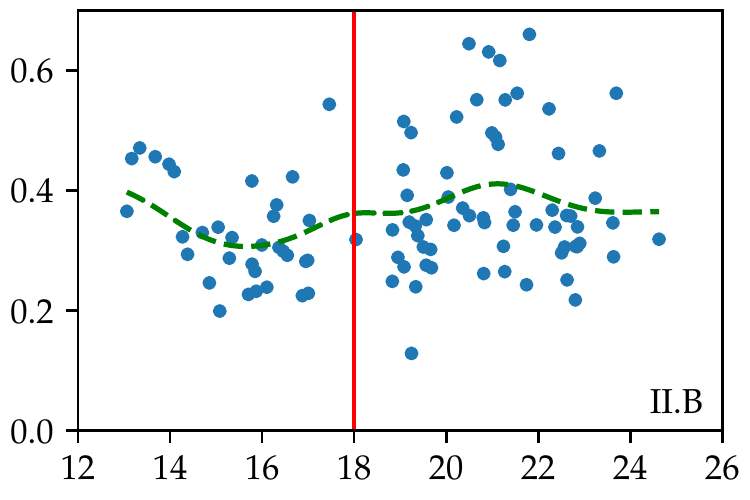}
		\end{minipage} &
		\begin{minipage}{\width}
			\includegraphics[width=\linewidth]{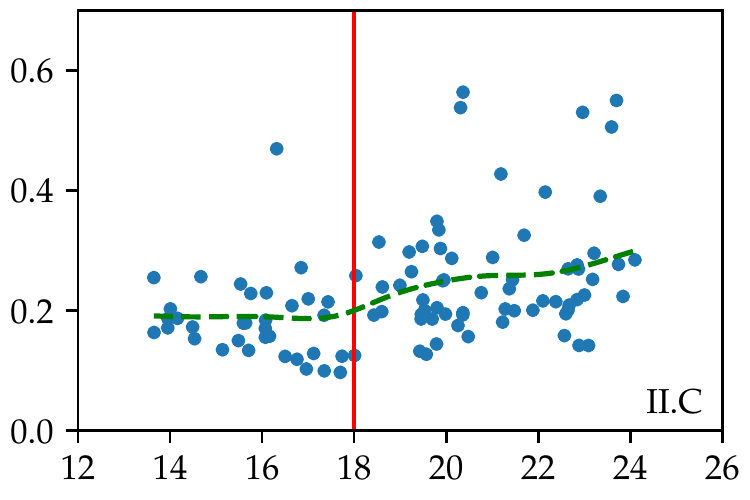}
		\end{minipage} \\
		& \rotatebox{90}{\makebox[0em]{~~~~~\tiny{Aleatoric Uncertainty}}} &
		\begin{minipage}{\width}
			\includegraphics[width=\linewidth]{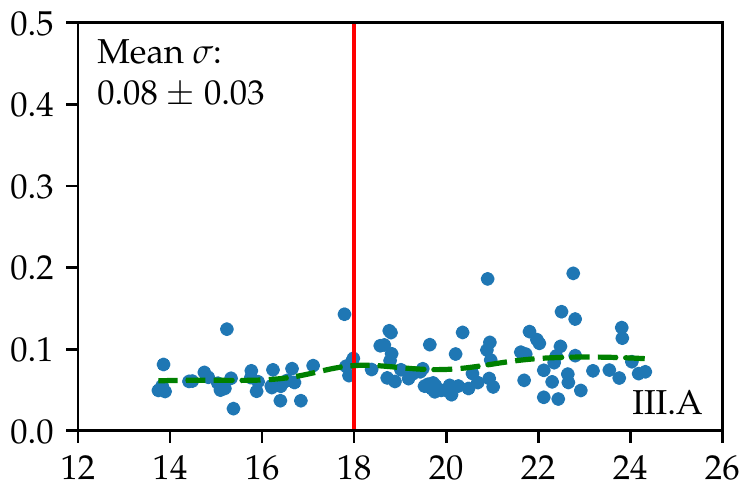}
		\end{minipage} & &
		\begin{minipage}{\width}
			\includegraphics[width=\linewidth]{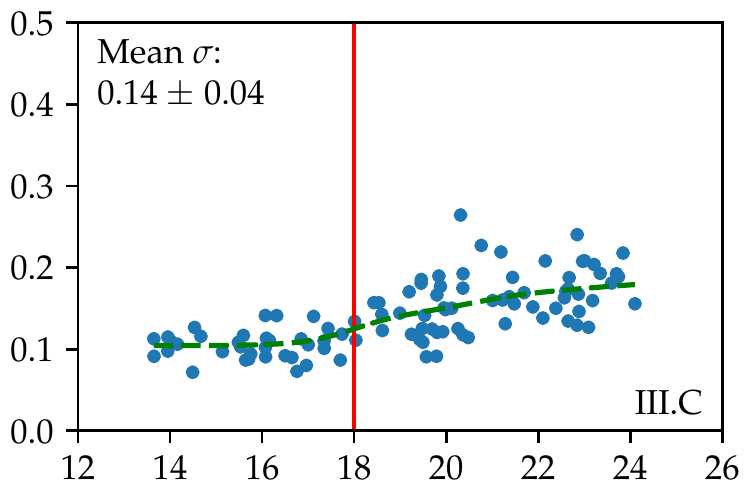}
		\end{minipage}\\
		\midrule
		\multirow{2}{*}{\rotatebox{90}{\makebox[2.5cm]{\small Trained on Wrist Bones}}} & \rotatebox{90}{\makebox[0em]{~~~~~\tiny{Epistemic Uncertainty}}} & &
		\begin{minipage}{\width}
			\includegraphics[width=\linewidth]{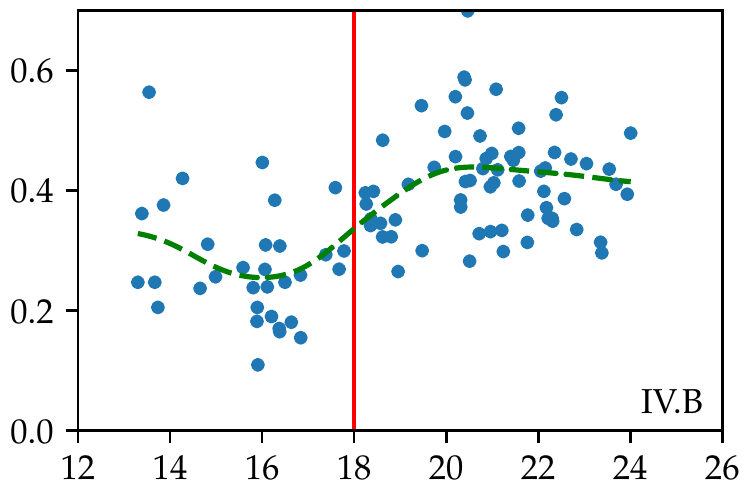}
		\end{minipage} &
		\begin{minipage}{\width}
			\includegraphics[width=\linewidth]{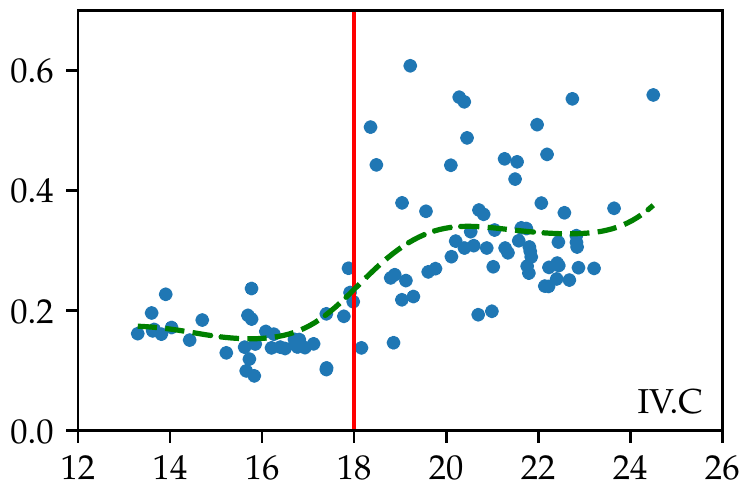}
		\end{minipage} \\
		& \rotatebox{90}{\makebox[0em]{~~~~~\tiny{Aleatoric Uncertainty}}} &
		\begin{minipage}{\width}
			\includegraphics[width=\linewidth]{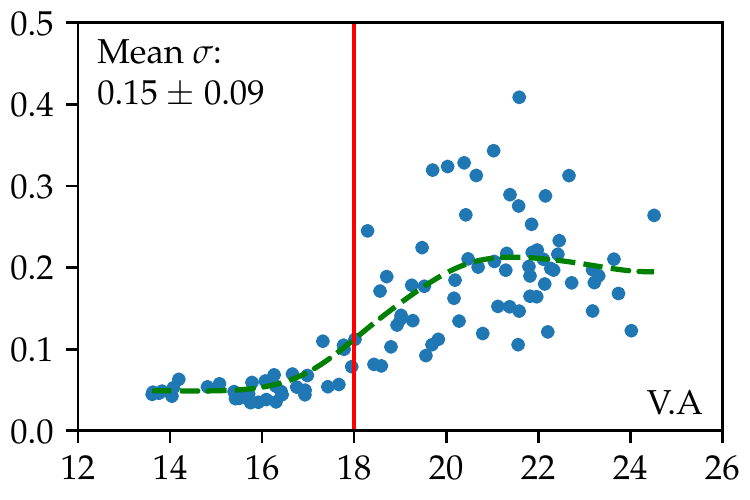}
		\end{minipage} & &
		\begin{minipage}{\width}
			\includegraphics[width=\linewidth]{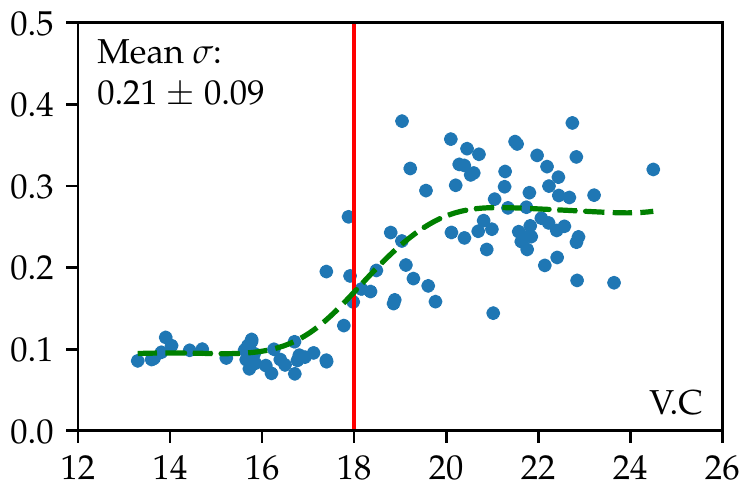}
		\end{minipage}\\
		& & \tiny{Predicted Age} & \tiny{Predicted Age} & \tiny{Predicted Age}\\
	\end{tabular}
	\caption{Scatter plots of predicted and groundtruth age, as well as predicted age with aleatoric and epistemic uncertainties for both networks trained on wrist and clavicle bones, and wrist bones alone.}
	\label{fig:plots}
\end{figure}

In forensic age estimation we are interested in both accurate BA prediction and the range of CA of subjects with the same BA.
Measured in mean absolute deviation between predicted BA and CA of a subject, the accuracy of the probabilistic BCNN (BCNN with~$\sigma$, $1.04 \pm 0.88$~y) is higher, compared to the standard CNN ($1.12 \pm 0.93$~y).
Despite using only 2D middle slices of wrist and clavicle bones, this result can also be considered in-line with the state-of-the-art method for multi-factorial age estimation in 3D MR images \cite{Stern2018}. 
Although a distribution placed over the weights allows capturing both epistemic and aleatoric uncertainty, in this work we are separating them by modeling uncertainty of the data with a Gaussian layer at the output of the network.
When comparing a plain BCNN and our BCNN with a Gaussian output layer, we see that data uncertainty learned at the output of the network improves the accuracy (${1.15 \pm 0.88}$ vs. ${1.04 \pm 0.88}$~y).

The main challenge in evaluating the quality of predictive uncertainty is the unavailable "ground truth" uncertainty. 
However, specific to the task of age estimation, biological variation, i.e. subjects with different CA having the same BA, can be interpreted as uncertainty of the prediction. 
This can best be seen when comparing aleatoric uncertainty of the multi-factorial BA estimation (Fig.~\ref{fig:plots} III.C) with BA estimation from wrist bones only (Fig.~\ref{fig:plots} V.C) for the subjects older than $18$ years.
Namely, due to saturation of aging information in the wrist images after the age of $18$ years, the model is not capable of distinguishing the subjects' age without including clavicle bones in the age prediction.
Thus, aleatoric uncertainty of the network rapidly increases after the age of $18$ when the network is trained with wrist bones only (Fig.~\ref{fig:plots} V.C). 
Although the same behavior can be seen in the CNN combined with the aleatoric uncertainty (Fig.~\ref{fig:plots} V.A), it is more prominent in our BCNN with separated data and model uncertainty (Fig.~\ref{fig:plots} V.C). 
Moreover, without separating these two uncertainties, the increase in the uncertainty of BCNN models trained solely from wrist bones is not clearly visible (Fig.~\ref{fig:plots} IV.B).
Finally, the small increase in the aleatoric uncertainty can also be seen when our BCNN is trained with wrist and clavicle bones (Fig.~\ref{fig:plots} III.C), which can be interpreted as being due to the less distinguishable ossification progress in clavicle bones compared to wrist bones. 
In our future work we will compare our VI based BCNN method with the approach proposed in \cite{Gal2016} that uses dropout during inference to approximate a Bayesian network.
In conclusion, compared to a standard CNN, we have shown that a Bayesian CNN not only provides more accurate results, but is also able to estimate the uncertainty of multi-factorial age estimation from MRI data.

\bibliographystyle{plain}
{\footnotesize\bibliography{bibliography/library.bib}}

\end{document}